\documentclass[twocolumn,prl,10pt,superscriptaddress]{revtex4-1}
\usepackage[latin1]{inputenc}
\usepackage[dvips]{graphicx}

\usepackage{dcolumn}
\usepackage{bm}
\usepackage{dsfont}
\usepackage{color}
\usepackage{url}
\usepackage[colorlinks=true ,citecolor=blue]{hyperref}
\usepackage{amsmath,amssymb,esint}
 \usepackage[table]{xcolor}

\definecolor{Nathanblue}{rgb}{0.,0.24,0.51}

\newcommand{\be}{\begin{equation}}
	\newcommand{\ee}{\end{equation}}
\newcommand{\bq}{\begin{eqnarray}}
	\newcommand{\eq}{\end{eqnarray}}

\newcommand{\1}{1\!\!1}

\newcommand\beq{\begin{equation}}
\newcommand\eeq{\end{equation}}
\newcommand\bal{\begin{aligned}}
\newcommand\eal{\end{aligned}}

\newcommand\snw{\mathfrak{snw}}

\begin{document}

\title{Non-Relativistic Supergeometry in the Moore\,-Read Fractional Quantum Hall State}

\author{Patricio Salgado-Rebolledo}
\affiliation{Department of Theoretical Physics, Wroc\l{}aw University of Science and Technology, 50-370 Wroc\l{}aw, Poland}
\affiliation{Universit\'e Libre de Bruxelles and International Solvay Institutes, ULB-Campus Plaine CP231, B-1050 Brussels, Belgium}

\author{Giandomenico Palumbo}
\affiliation{School of Theoretical Physics, Dublin Institute for Advanced Studies, 10 Burlington Road,
	Dublin 4, Ireland}

\date{\today}

\begin{abstract}
\noindent The Moore-Read state is one of the most well known non-Abelian fractional quantum Hall states.
It supports non-Abelian Ising anyons in the bulk and a chiral bosonic and chiral Majorana modes on the boundary. It has been recently conjectured that these modes are superpartners of each other and described by a supersymmetric conformal field theory \cite{Yang}. We propose a non-relativistic supergeometric theory that is compatible with this picture and gives rise to an effective spin-3/2 field in the bulk. After breaking supersymmetry through a Goldstino, the spin-3/2 field becomes massive and can be seen as the neutral collective mode that characterizes the Moore-Read state. By integrating out this fermion field, we obtain a purely bosonic topological action that properly encodes the Hall conductivity, Hall viscosity and gravitational anomaly. Our work paves the way to the exploration of the fractional quantum Hall effect through non-relativistic supergeometry.
\end{abstract}

\maketitle
\noindent {\bf Introduction: }
The Fractional Quantum Hall Effect (FQHE) is a milestone in the understanding of topological phases of matter \cite{Wen}. Non-Abelian FQH states have been investigated since the seminal work by Moore and Read \cite{Moore-Read}.
The Moore-Read (MR) state is a candidate to describe the filling fraction $5/2$ and represents the prototypical example of FQH state that supports non-Abelian anyons in the bulk with a single chiral boson and chiral Majorana mode propagating on the boundary. In the low-energy regime, these edge modes are described by a $SU(2)_2$ Wess-Zumino-Witten (WZW) model with chiral central charge $c=1+1/2$, which is associated to a $SU(2)_2$ Chern-Simons (CS) theory in the bulk \cite{Fradkin3,Palumbo5,Palumbo6}.
However, recently it has been proposed an alternative picture of the boundary theory of the MR state, in which the fermionic and bosonic modes can be seen as superpartners such that the effective theory is given by a chiral supersymmetric conformal field theory (SCFT) \cite{Yang, Bae, Sagi}.
The emergence of supersymmetry in this framework is partially justified by the possible existence of a propagating massive spin-3/2 mode in the bulk \cite{Bonderson,Moller,Yang2}. This collective mode in the MR state has been argued to indeed be the superpartner \cite{Yang2,Gromov} of the spin-2 GMP (Girvin-MacDonald-Plazman) mode \cite{GMP,Cappelli2, Karabali}. The latter can be seen a non-relativistic massive graviton \cite{Golkar,Liu,Liou,Gromov2,Gromov3} while the former would be a massive gravitino (gravitinos carry spin-3/2 in supergravity). We remind here that without being related to the MR state, supersymmetry has been previously discussed in a number of works concerning the FQHE \cite{Tong,Vafa,Ino,Hasebe,Hik,Fuji}.

In this framework, it is natural to expect that the topological sector of the supersymmetric model describing the MR state is given by a (2+1)-dimensional CS theory invariant under some non-relativistic superalgebra similar to three-dimensional non-relativistic supersymmetry or supergravity \cite{Horvathy:1992pcm,Duval:1993hs,Andringa:2013mma,Bergshoeff:2016lwr,Ozdemir:2019tby,Concha:2020tqx}. If this supersymmetric picture of the MR state is correct, then one should also be able to take into account further geometric effects such as the Hall viscosity \cite{Avron,Read,Read-Rezayi,Bradlyn4,Hoyos-Son}, which is an universal feature of the FQHE and emerges once the FQH fluid is coupled to a curved background. Several works have discussed the geometric effective non-relativistic field theory of the Abelian FQH states \cite{Gromov2,Bradlyn,Fradkin,Fradkin2,Cappelli, Bradlyn3,Gromov-Jensen,Moroz,Wiegmann,Golan-Hoyos,Gromov-Abanov,Abanov-Gromov,Geracie2,Wiegmann2,Ferrari,Dwivedi}, where the effective action is purely bosonic as well as its corresponding edge states.

The main goal of our work is to fill the gap between supersymmetry and the geometric theory for the MR state, by introducing and studying a novel non-relativistic supergeometric theory that is able to describe both the chiral boundary states and the topological and geometric bulk features, such as Hall viscosity and the presence of massive spin-3/2 mode. In order to combine supersymmetry with a geometric theory compatible with the symmetries of the MR state, we start by defining a novel supersymmetric extension of the extended Nappi-Witten algebra, which has been introduced in an our previous work that deals with the geometry of the Abelian Laughlin state \cite{Salgado-Rebolledo:2021wtf}. 
Through this new algebra, we will define the corresponding gauge connections that contains both the external and hydrodynamical fields. The universal topological action is given by the CS term associated to this connection. However, in order to obtain a massive spin-3/2 mode we will add an extra gauge-invariant term in the action \cite{Palumbo4} that introduces a Goldstino field and breaks supersymmetry, giving rise to the mass term of the spin-3/2 field. This will allow us to integrate our the fermionic field in the bulk and obtain a full bosonic geometric theory that encodes the Hall viscosity response of the system. On the boundary, the effect of adding this term is that the chiral boson and chiral Majorana modes acquire different velocities, breaking supersymmetry at the level of the edge theory as well.

Our work offers a non-relativistic supergeometric framework to describe the universal (topological) features of the MR state in the low-energy regime. It represents a novel bridge between two apparently uncorrelated research fields, i.e. non-relativistic supersymmetry and FQHE. Although our supersymmetric theory is formally different with respect to other non-relativistic supergravity models presented previously in the high-energy-physics context, it is compatible with the underlying symmetries of the FQHE and paves the way for the construction of a more general theory describing dynamical (non-universal) features of massive gravitons and gravitinos.

\noindent {\bf Nappi-Witten algebra and unbroken supersymmetry: }
In order to build the proper supergeometric theory for the Moore-Read state, we need first to specify the underlying algebra compatible with the symmetries of the FQHE. In fact, it is well known that general relativity, in the first-order formalism, can be seen an effective gauge theory associated to the Lorentz group. In this formalism the metric tensor is replaced by the spin connection and the vielbein. In the (2+1)-dimensional case, both exotic (see, Ref.\cite{Palumbo3} for an application of exotic gravity in topological phases) and Einstein-Hilbert gravity, with and without cosmological constant, can be rewritten as a CS theory where the corresponding gauge connection is a linear combination of the spin connection and the dreibein (i.e. the equivalent of the vielbein in 2+1 dimensions) \cite{Witten,Achucarro}. Differently from these relativistic theories, an effective geometric action for FQHE should take into account the magnetic translations together with the Galilei invariance. The latter, although not crucial in the FQHE, has been shown to be very useful in the construction of non-relativistic effective theories in several FQH states. In our previous work, we have shown that both magnetic translations and space rotations are naturally encoded in an extended version of the Nappi-Witten algebra \cite{Nappi-Witten,Figueroa-OFarrill:1994liu}, which allowed us to build a non-relativistic geometric theory for the Laughlin state \cite{Salgado-Rebolledo:2021wtf}. By following Ref.\cite{Yang}, we consider now, besides the previously mentioned symmetries, also supersymmetry as a symmetry for the Moore-Read state (in the unbroken supersymmetric phase). In this way, by identifying the correct supersymmetric generalization of the extended Nappi-Witten algebra we will be able to construct a corresponding non-relativistic supergeometric theory by gauging the global symmetries. This non-relativistic supersymmetric algebra exists and is given by
\beq \label{SNW}
\bal
&
\left[ \mathcal  J, \mathcal  P_a\right]=\epsilon_a^{\;\; b} \mathcal P_b, &\,&
\left[ \mathcal  P_a,  \mathcal P_b\right]=-\epsilon_{ab} \mathcal T ,
\\
&
\left[ \mathcal  J , \mathcal  Q_\alpha \right]=\frac{1}{2} \left(\gamma_0 \right)_{\;\;\alpha}^{\beta}  \mathcal Q_\beta,
&\,&
\left\{  \mathcal Q_\alpha,  \mathcal Q_\beta\right\}= \frac{1}{2}\left(C \gamma_0 \right)_{\alpha\beta} \mathcal T,
\\&
\left[ \mathcal  Z, \mathcal  P_a\right]=\beta\epsilon_a^{\;\; b} \mathcal P_b, &\,&
\left[ \mathcal  Z , \mathcal  Q_\alpha \right]=\frac{\beta}{2} \left(\gamma_0 \right)_{\;\;\alpha}^{\beta}  \mathcal Q_\beta.
\eal
\eeq
where $\beta$ is an arbitrary constant, $C_{\alpha\beta}=\epsilon_{\alpha\beta}$ is the charge conjugation matrix and $\gamma_0=i\sigma_2 $. In the bosonic sector $\mathcal  P_a$ ($a=1,2$), stands for the generator of translations in the two-dimensional plane, $\mathcal J$ is the generator of the spatial rotations, $\mathcal T$ is the Lie algebra generator associated to the external electromagnetic field and $\mathcal Z$ is a bosonic generator associated to the emergent U(1) gauge field. The fermionic generators $Q_\alpha$ ($\alpha=1,2$) are Majorana supercharges. The first line in \eqref{SNW} represents the Euclidean version of the original algebra by Nappi and Witten \cite{Nappi-Witten}, which is ubiquitous in low-dimensional physics \cite{Alvarez:2007ys,Afshar:2019axx,Duval:2000xr,Horvathy:2004am,Palumbo2} and is also related to the Maxwell algebra \cite{Schrader:1972zd,Gomis:2017cmt} in the sense that it includes magnetic translations. The case $\beta=0$ yields a supersymmetric extension of the Nappi-Witten symmetry. When the supercharges $Q_\alpha$ vanish, we recover the extended Nappi-Witten algebra employed for the Laughlin state in \cite{Salgado-Rebolledo:2021wtf}. From now on, we will refer to the symmetry \eqref{SNW} as $\snw$ algebra.

From this novel supersymmetric algebra, by following the gauge principle, we can naturally build a corresponding one-form gauge connection
\beq\label{connection}
\mathbb A= \omega \mathcal J+a \mathcal Z+e^a \mathcal  P_a   + A \mathcal  T  + \mathcal Q_\alpha \Psi^\alpha,
\eeq
where $\omega$ is the spin connection, $a$ is an emergent gauge field, $e^a$ is the spatial dreibein, $A$ is the external electromagnetic field, and $\Psi^\alpha$ is a fermionic spin-3/2 field.
They are all the physical fields in the effective geometric theory that we will construct in the next section. Note that our superalgebra includes the central generator $\mathcal T$ in the anti-commutator of the supercharges, in analogy with the super-Bargmann algebra \cite{Puzalowski:1978rv,deAzcarraga:1991fa,Bergman:1995zr}. Since the Hamiltonian generator does not appear in the $\{Q,Q\}$ bracket, this algebra does not lead to supersymmetry in the standard sense. In order to include the Hamiltonian generator, it is necessary to embed the algebra \eqref{SNW} in a lager non-relativistic superalgebra, as done in \cite{Salgado-Rebolledo:2021wtf} in the purely bosonic case. In the same way, the supersymmetric generators cannot be understood as the square root of the spatial translations $\mathcal P_a$, as it happens in non-relativistic supergravity \cite{Andringa:2013mma,Bergshoeff:2016lwr,Ozdemir:2019tby,Concha:2020tqx} or superparticle models \cite{Lukierski:2006tr,Bergshoeff:2014gja}. Thus, the theory that will follow from gauging $\snw$ will not define a supergravity theory.
\\

\noindent {\bf Supergeometric Chern-Simons theory and boundary chiral modes: }
In previous section, we have introduced a novel algebra $\snw$, defined by the commutation relations \eqref{SNW}, that encodes two-dimensional magnetic translations, spatial rotations and supersymmetry. Another fundamental feature in the FQHE is the absence of the time-reversal symmetry, which is typically taken into account by considering a CS theory, which is a topological field theory that supports a CFT on the boundary of the system. Because we are mainly interested in the topological (universal) features of the MR state in the low-energy regime, a natural choice for the construction of an effective field theory in terms of the gauge connection \eqref{connection} is to consider a CS action of the form 
\beq\label{CSaction}
S_{\rm CS}=-\frac{k}{4\pi}\int_{M} \left\langle\mathbb A d \mathbb A+ \dfrac{2}{3} \mathbb A \wedge \mathbb A  \wedge \mathbb A \right\rangle.
\eeq
where $M$ is a three-dimensional spacetime manifold and $k$ is the corresponding quantized level, which we fix $k=1$ for simplicity.  Here, $\left\langle\cdot,\cdot \right\rangle$ denotes some non-degenerate invariant bilinear form on the gauge algebra \eqref{SNW}, i.e. denoting the generators of \eqref{SNW} by $\mathcal G_A=\{\mathcal J,\mathcal Z, \mathcal P_a, \mathcal T, \mathcal Q_\alpha\}$, with $A$ a collective index, the map $\left\langle\cdot,\cdot \right\rangle:\snw\times\snw\rightarrow \mathbb R$ satisfies
\beq\bal
&i)\;\left\langle \mathcal  G_A,\mathcal G_B \right\rangle=\left\langle \mathcal G_A +[\mathcal G_A,\mathcal G_C],\mathcal G_B + [\mathcal G_B,\mathcal G_C] \right\rangle,\\ &ii)\;{\rm det} \left\langle \mathcal G_A,\mathcal G_B \right\rangle\neq0.
\eal\eeq
Given an arbitrary gauge symmetry defined by some Lie algebra, the existence of such invariant form, and thus of a well-defined CS action, is not guaranteed. For example, the non-degenerate invariant bilinear form on the Poincar\'e algebra in three spacetime dimensions \cite{Witten:1988hc} is a peculiarity that does not occur in higher dimensions and allowed to define a gauge theory for three-dimensional Einstein gravity. Similarly, in the non-relativistic case, the Galilean symmetry does not admit a well-defined invariant form unless its double central extension is considered, which exists only in three space-time dimensions and allows to define a Bargmann-invariant non-relativistic CS gravity theory \cite{Papageorgiou:2009zc}. Along the same lines, a remarkable property of the extended supersymmetric Nappi-Witten algebra $\snw$ is that it admits a non-degenerate invariant bilinear form, which reads
\beq\label{invtensor}
\bal
&
\langle \mathcal J, \mathcal J \rangle=\mu_0\,, 
\qquad
\langle \mathcal P _a , \mathcal P  _b\rangle=\mu_1\,\delta_{ab}\,,
\\[5pt]&
\langle \mathcal J , \mathcal Z \rangle=\mu_2,
\qquad
\langle \mathcal Q_\alpha , \mathcal Q_\beta \rangle=\mu_1\,C_{\alpha\beta}, 
\\[5pt]&
\langle \mathcal Z, \mathcal Z \rangle= \mu_3,
\qquad
\langle  \mathcal J , \mathcal T \rangle=
\beta^{-1}\langle  \mathcal Z , \mathcal T \rangle=- \mu_1,
 \eal
\eeq
where $\mu_i$ ($i=0,1,2,3$) are all arbitrary real parameters. Therefore, it is possible to define a well-defined CS action \eqref{CSaction} invariant under gauge transformations $\delta_\epsilon A= dA+[A,\epsilon]$ with $\epsilon$ a local parameter taking values in $\snw$. The construction of the CS action invariant under the symmetry \eqref{SNW} is the first result of our paper. As we will see in the next sections, such action can be used as an effective action for the Moore-Read fractional quantum Hall states, provided we associate the constants $\mu_i$ to physical quantities, such as Hall conductivity, Hall viscosity and chiral central charge. 

Without matter currents, the corresponding equations of motions are given by $\mathbb F=0$, where $\mathbb F$ is curvature tensor of $\mathbb A$, namely
\beq
\mathbb F= d \mathbb A +\frac{1}{2} [\mathbb A, \mathbb A] = d\omega \mathcal J+ d a \mathcal Z + R^a \mathcal P_a + F \mathcal T +  \mathcal Q_\alpha D \Psi^\alpha,
\eeq
where
\beq
\bal
&R^a = de^a + \epsilon^a_{\;\;b}  e^b \wedge\left( \omega + \beta a\right)\,,\\
&F= dA-\frac{1}{2}\epsilon_{ab}e^a \wedge e^b -\frac{1}{4}  \bar \Psi_\alpha \wedge  \left(\gamma_0 \right)^\alpha_{\;\;\beta}  \Psi^\beta,  \\
\eal
\eeq
and we have defined the covariant derivative of $\Psi^\alpha$ as
\beq
D\Psi^\alpha = d\Psi^\alpha + \frac{1}{2}  \left( \omega + \beta a\right) \wedge \left(\gamma_0\right)_{\;\;\beta}
^\alpha \Psi^\beta.
\eeq
By employing the above expressions and defining the conjugate spinor $\bar\Psi_\alpha=\Psi^\beta C_{\beta\alpha}$, we can rewrite the CS action in Eq.\eqref{CSaction} in terms of the physical fields
\beq\label{barggrav2}
\bal
&S_{\rm CS}=-\displaystyle\frac{1}{4\pi}\int_{M} \bigg[
\mu_0\,\omega\wedge d\omega   
+2\mu_2\, a \wedge d\omega 
+\mu_3\, a \wedge da
\\&+\mu_1\, e^a\wedge  R_a -2\mu_1\, A \wedge d\left(\omega + \beta  a\right)
-\mu_1\, \bar \Psi_\alpha \wedge D\Psi^\alpha  \bigg]\,.
\eal
\eeq
This a gauge invariant topological action for a manifold without boundary and will give rise to the topological response of the fractional Hall fluid in presence of the external fields as we will discuss in full detail in the next section.
We now move our attention to the case where $M$ has a boundary and consider the dynamical quantum modes induced by this effective topological action on the spatial boundary of the system. In fact, on a manifold with boundary a CS action is not gauge invariant anymore and dynamical gapless degrees of freedom need to appear on the boundary in order to compensate the gauge anomaly of the bulk  \cite{Dijkgraaf}.
This is the well known CS/CFT correspondence that holds also in our case as we now show. We start considering for simplicity a manifold $M= D_2 \times R$, where $R$ is associated to the time-like coordinate $t$, while $D_2$ is a two-dimensional disk on which we introduce polar coordinates $(r, \phi)$, such that its boundary is given by one-dimensional circle $S^1$. Thus, the boundary theory is given by the following chiral super-WZW model \cite{Coussaert:1995zp} for an element $g$ of the Lie group associated to the $\snw$ algebra,
\beq\label{wzwaction}
\bal
S_{\rm WZW}&=
\frac{1}{4\pi} \int dt d\phi \, \left\langle g^{-1}\partial_+  gg^{-1} \partial_\phi g \right\rangle \\
&+\frac{1}{12\pi}\int_{M}\left\langle\left(\tilde g^{-1} d\tilde g\right)^3\right\rangle,
\eal
\eeq
which is obtained after replacing the local solution of the CS constraint $\mathbb F_{ij}=0$ ($i=\{r,\phi\}$),
given by $\mathbb A_i= g^{-1} \partial_i g$ with the boundary condition $\left(\mathbb A_t + v \mathbb A_\phi\right)|_{\partial M}=0$
back in the action \eqref{CSaction}. Here, we consider a boundary group element that depends only on the boundary coordinates, i.e. $g=g(t,\phi)$. The group element $\tilde g$ in \eqref{wzwaction} is the extension of $g$ to the three-dimensional bulk $M$. In the gauge $\partial_r \mathbb A_\phi =0$, one finds
\beq\label{gh}
\tilde g(t,r,\phi) = g(t,\phi) h(t,r),
\eeq
for some $h(t,r)$. The result \eqref{wzwaction} then follows from introducing the right boundary term in the CS action such that $\delta S|_{\rm on-shell}=0$ and considering $\partial_t h|_{\partial M}=0$. Notice that $v$ will represent the velocity of the chiral modes, and we have defined $x^\pm = (1/2)\left(t\pm(1/v)\phi\right)$ and $ \partial_\pm= \partial_t \pm v\partial_\phi$. In order to find the explicit form of $S_{\rm WZW}$ we look at the left-invariant Maurer-Cartan form on $\snw$, $\Omega=g^{-1}dg$, which satisfies the Maurer-Cartan equation $
d\Omega +\Omega\wedge\Omega=0$. From \eqref{gh}, one can see that the bulk and boundary left-invariant Maurer Cartan forms are related by $\tilde\Omega= h^{-1} \Omega h + h^{-1} d h$. By employing the $\snw$ commutation relations \eqref{SNW}, the Maurer-Cartan equation can be solved and leads to
\beq
\Omega= d\theta\mathcal J  +d\varphi \mathcal Z+ \Omega_{\mathcal P}^a \mathcal P_a +\Omega_{\mathcal T}\mathcal T+  \mathcal Q_\alpha \Omega_{\mathcal Q}^\alpha,
\eeq
where we have defined
\beq
\bal
\Omega_{\mathcal P}^a&= d \sigma^a-\epsilon^a_{\;\;b}\sigma^b (d\theta+\beta d \varphi),\\
\Omega_{\mathcal T}&= d \vartheta+ \frac{1}{2}\epsilon_{ab} \sigma^a \Omega_{\mathcal P}^b- \frac{1}{4}\bar\chi_\alpha  \left(\gamma_0\right)^\alpha_{\;\;\beta}d\chi^\beta,\\
\Omega_{\mathcal Q}^\alpha&=\exp[-(1/2)\left(\theta+ \beta\varphi\right)\left(\gamma_{0}\right)^\alpha_{\;\;\beta}] d\chi^\beta,\\
\eal
\eeq
where $\chi^\alpha$ is a Grassmann-valued spinor. The one-form $\tilde \Omega$ has the same functional form as $\Omega$ in terms of fields $\tilde\theta$, $\tilde\varphi$, $\tilde\sigma^a$ and $\tilde \chi$ that include $r$-dependent functions which properly decouple when expressing the Wess-Zumino term in \eqref{wzwaction} as a boundary integral.

By putting now all these solutions back into $S_{\rm WZW}$, one finds that $\vartheta$ is the Lagrange multiplier that enforces the constraint
$\partial_+  \theta^\prime + \beta \partial_+  \varphi^\prime =0$, which implies
$\theta + \beta\varphi = \rho(t) + \lambda(x^-)$, with $\rho(t)$ and $\lambda(x^-)$ arbitrary functions of their arguments. Replacing this expression in the action and integrating out the field $\sigma^a$, we find
\beq\label{wzwaction4}
\bal
S_{\rm WZW}
&=\frac{1}{4\pi}\int dt d\phi \Big(\tilde{\mu}\,\partial_\phi \varphi \partial_+ \varphi +\mu_1  \psi \partial_+ \psi
\Big),
\eal
\eeq
where we have imposed the extra fermionic boundary condition $\chi^1|_{\partial M}={\rm constant}$ (inspired in the three-dimensional supergravity analysis \cite{Henneaux:1999ib,Barnich:2015sca}), introduced the new field $\psi= (\sqrt{v\partial_- \lambda }/2) \chi^2$, and defined $\tilde{\mu}=\mu_3 -2\beta\mu_2+\beta^2 \mu_0$. This chiral CFT describes the chiral boson and chiral Majorana modes that propagate on the boundary of the system in agreement with the boundary states of the MR state. This represents one of the main results of our work. Notice that in the unbroken supersymmetric limit, both modes have the same velocity \cite{Yang}. We will indeed analyze the broken supersymmetric phase of our theory in the next section, where we will introduce a Goldstino field that induces a mass term to the spin-3/2 fermion.\\

\noindent {\bf Broken supersymmetry and topological response: }
Supersymmetry breaking can be obtained by introducing a fermionic Goldstone field, known as Goldstino \cite{Volkov}. Importantly, this new field induces a mass term for the spin-3/2 field \cite{Deser-Zumino,Dereli,Deser}, which is expected to be a massive collective mode of the bulk.
Here we adopt the approach proposed in Ref.\cite{Palumbo4} by one of the authors, where it has been shown that there exists a non-propagating fermionic field in 2+1 dimensions that is able to induce a spin-3/2 mass term. We interpret this fermion as a Goldstino within our current framework, described by the action
\beq\label{matter}
S_{\rm \eta}=\frac{\mu_1}{4 \pi}\int_{M} d\bar{\eta}_\alpha \wedge \hat{\gamma}^\alpha_{\;\;\beta} \wedge \,d\eta^\alpha,
\eeq
where $\eta^\alpha$ is the Goldstino spinor field describing a neutral fermion, and $\bar{\eta}_\alpha=\eta^\beta C_{\beta\alpha}$ is its conjugate. We have also introduced the matrix $\hat{\gamma}^\alpha_{\;\;\beta}=\tau \left(\gamma_{0}\right)^\alpha_{\;\;\beta}$, where $\tau$ is some fixed non-dynamical clock form defining a Newton-Cartan structure. The term \eqref{matter} is therefore a boundary term and does not contribute to the bulk dynamics of the system. This action is metric independent and so far does not introduce further propagating degrees of freedom in our theory. Importantly, it partially resembles the higher-derivative term that appears in the Volkov-Akulov action for the relativistic Goldstino in 2+1 dimensions \cite{Sorokin}. We can naturally couple the Goldstino to the spin-3/2 field as follows
\beq
d\eta^\alpha\rightarrow d\eta^\alpha-\sqrt{m}\Psi^\alpha, \hspace{0.4cm} d \bar{\eta}_\alpha\rightarrow d\bar{\eta}_\alpha-\sqrt{m}\bar{\Psi}_\alpha,
\eeq
with $m$ a dimensionful parameter. This real coefficient is indeed the mass of the spin-3/2 field. In fact, by replacing the above expressions in $S_{\eta}$ we obtain the following term
\beq\label{mass}
S_{m}=\frac{\mu_1\,m}{4 \pi}\int_{M} \bar{\Psi}_\alpha\wedge \hat{\gamma}^\alpha_{\;\;\beta}\wedge \Psi^\beta,
\eeq
which the simplest mass term that can be built for the spin-3/2 field and it is also parity-odd similarly to the Dirac mass in the spin-1/2 theory \cite{Dereli,Deser}. This term breaks local supersymmetry and is compatible with the possible existence of a massive spin-3/2 collective mode. We can now integrate out $\Psi^\alpha$ in the bulk action $S_{\rm CS}+S_m$ to obtain a purely bosonic effective action that will allow us to derive the topological response of the MR state.
Before integrating out $\Psi^\alpha$, we rescale it $\Psi^\alpha\rightarrow (\sqrt{8\pi})\Psi^\alpha$ and by assuming that $m$ is large and positive and neglecting the terms that couple $\eta^\alpha$ and $\Psi^\alpha$ (they are proportional to $\sqrt{m}$, which we consider small compared to $m$), we finally obtain a CS term $(\mu_1 \varepsilon/4\pi) (\omega+ \beta a)\wedge d(\omega+\beta a)$ (with $\varepsilon={\rm sign}(m)$) that adds up to the other CS terms in \eqref{barggrav2} \cite{Gaume,Grimm}. 
Moreover, in order to have a unique geometric response from the background geometry, we vary the action with respect to the spatial dreibein $e^a$ to obtain the field equation $R_a=0$, which in turn yields the following equation for the torsion
\beq \label{torsion}
T^a \equiv d e^a +\epsilon^a_{\;\;b} \wedge e^b \omega= -\beta\epsilon^a_{\;\;b} e^b \wedge a.
\eeq
This equation allows us to formally express the dreibein in terms of the spin connection and the field $a$.
Thus, the effective action that depends only on the external fields is obtained by integrating out also the spin-1 hydrodynamic field $a$ such that
\begin{eqnarray}\label{barggrav3}
S[A, \omega]=\displaystyle\frac{1}{4\pi}\int_{M} \Bigg[
\hat{c}\,\omega\wedge d\omega    
+\nu A \wedge d \left(A+
 2\bar s \,\omega\right) \Bigg].
\end{eqnarray}
Here, for the MR state, $\hat{c}=\nu\bar s^2-c/12$, where $\nu=1/2$, $\bar s=c=3/2$ are the filling factor, average orbital spin and chiral central charge, respectively \cite{Fradkin,Gromov2}. The first and third terms in the above action are known as gravitational Chern-Simons \cite{Fradkin} and Wen-Zee term \cite{Wen-Zee}, respectively. The former is associated to the gravitational anomaly, while the latter is related to the Hall viscosity $\eta_H$ \cite{Avron, Read,Read-Rezayi,Hoyos-Son,Bradlyn4}, which represents the response of the Hall fluid to shear or strain \cite{Geracie,Bradlyn}.

The physical coefficients in the action are obtained by fixing the arbitrary constants $\mu_i$ in the invariant bilinear form \eqref{invtensor} as $\mu_0=2\varepsilon\nu\bar s+c/12$, $\mu_1=2\nu\bar s$ and $\mu_2=2\beta\nu\bar s(\varepsilon+\bar s)$ and $\mu_3=2\beta^2\nu\bar s(\varepsilon+2\bar s)$ in our theory with $\varepsilon=1$. Note that the results obtained in \cite{Salgado-Rebolledo:2021wtf} for Laughlin states can be recovered by setting $\varepsilon=0$ (which is compatible with removing the fermion field $\Psi^\alpha\rightarrow 0$) together with $\nu=1/k$, $\bar s=k/2$, $c=1$ and a particular choice of $\beta$.

Introducing the Goldstino in the system does not only break supersymmetry in the bulk, but also at the boundary. Indeed, considering $\tau=dt$ and the boundary value of spin-3/2 field given by the Maurer-Cartan form $\Omega_{\mathcal Q}^\alpha$, one finds that the contribution of the term \eqref{mass} to the boundary action reads
\beq\label{shiftv}
\frac{ \mu_1m v \partial_{-}\lambda}{16 \pi}\int dt d\phi \;  \psi \partial_\phi \psi.
\eeq
where, without loss of generality we have assumed $\partial_{-}\lambda$ to be an arbitrary constant. Thus, the effect of the Goldstino at the boundary is a shift in the velocity of the chiral Majorana fermion $v\rightarrow v'= v+ mv \partial_{-}\lambda/4$, breaking the supersymmetry of \eqref{wzwaction4}. Note that $\beta$ is still arbitrary and, by choosing $\beta=\sqrt{\mu_1/\mu_0}$, the boundary theory obtained by adding \eqref{shiftv} to $S_{\rm WZW}$ takes the form 
\beq\label{bdytheory}
\bal
S_{\rm bdy}
&=\frac{\nu\bar s}{2\pi}\int dt d\phi \Big(\partial_\phi \varphi(\partial_t+v\partial_\phi) \varphi 
+\psi (\partial_t+v'\partial_\phi) \psi
\Big),
\eal
\eeq
compatible with the broken supersymmetric phase of the edge theory of the MR state \cite{Yang}.

\noindent {\bf Conclusions and outlook: }
Summarizing, in this paper we have proposed a novel geometric model for the Moore-Read FQH state based on non-relativistic supergeometry by introducing the novel gauge algebra $\snw$ and considering the corresponding CS action. Differently from canonical supergravity, our theory is not relativistic and naturally generalizes the non-supergeometric actions introduced in literature for the Abelian FQH states. On one hand, our topological action naturally encodes a spin-3/2 fermion in the bulk, which is expected to emerge as a collective mode in the MR state. On the other hand, our theory gives rise to a chiral supersymmetric CFT that decomposes into a chiral boson and a chiral Majorana fermion. Finally, we have shown that when supersymmetry is broken due to an effective Goldstino, the spin-3/2 mode acquires a mass and can be integrated out, providing the correct topological bosonic action that takes into account both the Hall conductivity and Hall viscosity. At the same time, supersymmetry is broken at the boundary, where the effect of adding the Goldstino is to modify velocity of the chiral Majorana field, recovering the result of \cite{Yang}. It would be very interesting to embed our supergeometric theory into a generalized super-bimetric model by following Ref.\cite{Gromov2} to include the massive spin-2 GMP mode as the superpartner of the spin-3/2 neutral fermion. Moreover, our supergeometric approach could be important for the geometrical characterization of hierarchies of FQH states in the second Landau level build from the MR state \cite{Bonderson2}. As it happens in the bosonic case \cite{Bergshoeff:2017vjg}, one could expect that the right massive wave equation for the spin-3/2 field in the FQHE can be obtained from a special non-relativistic limit of a spin-3/2 generalization of the Fierz-Pauli equation in 2+1 dimensions. It would be as well interesting to embedded our model into a full supersymmetric Newton-Cartan geometry, as done in \cite{Salgado-Rebolledo:2021wtf} in the bosonic case by means of Lie algebra expansions \cite{Hatsuda:2001pp,deAzcarraga:2002xi,Izaurieta:2006zz} following the method introduced in \cite{Penafiel:2019czp}.  We leave all these important open points to future work.

\noindent {\bf Acknowledgments: }
The authors are pleased to acknowledge discussions with Eric Bergshoeff, Andrea Cappelli, Carlos Hoyos. We thank Dung Nguyen for carefully reading the manuscript and for pointing out important details that led us to improve the article. P.S-R. has received funding from the Norwegian Financial Mechanism 2014-2021 via the National Science Centre (NCN) POLS grant 2020/37/K/ST3/03390. This work was partially supported by FNRS-Belgium (conventions FRFC PDRT.1025.14 and  IISN 4.4503.15), as well as by funds from the Solvay Family.

\appendix

%

\end{document}